\documentclass[twoside]{dis07}
\usepackage[latin1]{inputenc}
\usepackage[dvips]{graphicx,epsfig,color}
\usepackage{wrapfig,rotating}
\usepackage{amssymb,amsmath,array}

\begin{document}
\title{BFKL Effects in Azimuthal Angle Correlations of Forward Jets}

%***********************************************************************
% AUTHORS INFORMATION AREA
%***********************************************************************
\author{Agust{\' \i}n~Sabio~Vera$^1$ and Florian~Schwennsen$^2$
%
% Optional short acknowledgment: remove next line if non-needed
\thanks{Supported by the Graduiertenkolleg 
``Zuk\"unftige Entwicklungen in der Teilchenphysik''}
%
% DO NOT MODIFY THE FOLLOWING '\vspace' ARGUMENT
\vspace{.3cm}\\
%
% Addresses and institutions (remove "1- " in case of a single institution)
1- Physics Department - Theory Division CERN \\
CH--1211 Geneva 23 - Switzerland\\
%
% Remove the next three lines in case of a single institution
\vspace{.1cm}\\
2- II. Institut f\"{u}r Theoretische Physik - Universit\"{a}t Hamburg\\
Luruper Chaussee 149 D--22761 Hamburg -  Germany\\
}
%***********************************************************************
% END OF AUTHORS INFORMATION AREA
%***********************************************************************

\maketitle

\begin{abstract}
The azimuthal angle correlation of Mueller--Navelet jets at hadron colliders 
is studied in the NLO BFKL formalism. We highlight the need of collinear 
improvements in the kernel to obtain good convergence properties and we obtain 
better fits for the Tevatron data than at LO accuracy. We also estimate these 
correlations for larger rapidity differences available at the LHC.
\end{abstract}

\section{BFKL cross sections}
In~\cite{Vera:2007kn} we continue the study initiated in~\cite{Vera:2006un} of azimuthal 
correlations 
in Mueller--Navelet jets~\cite{Mueller:1986ey} using the Balitsky--Fadin--Kuraev--Lipatov (BFKL) 
equation in the next--to--leading (NLO) approximation~\cite{FLCC}. 
We investigate normalized differential cross sections which are quite insensitive to parton 
distribution functions and read 
\begin{eqnarray}
\frac{d {\hat \sigma}}{d^2\vec{q}_1 d^2\vec{q}_2} &=& \frac{\pi^2 {\bar \alpha}_s^2}{2} 
\frac{1}{q_1^2 q_2^2} \int \frac{d\omega}{2 \pi i} e^{\omega {\rm Y}} f_\omega \left(\vec{q}_1,\vec{q}_2\right),\nonumber
\end{eqnarray}
where ${\bar \alpha}_s = \alpha_s N_c / \pi$, $\vec{q}_{1,2}$ are the transverse momenta 
of the tagged jets, and Y their relative rapidity. The Green's function carries the 
Y--dependence and follows the NLO equation, $\left(\omega - {\bar \alpha}_s {\hat K}_0 
- {\bar \alpha}_s^2 {\hat K}_1\right) {\hat f}_\omega =  {\hat 1}$, 
which acts on the basis including the azimuthal angle, $\left< \vec{q}\right|
\left.\nu,n\right> = \frac{1}{\pi \sqrt{2}} \left(q^2\right)^{i \nu -\frac{1}{2}} 
\, e^{i n \theta}$. 
As Y increases the azimuthal dependence is driven by the kernel. This is why we use the LO 
jet vertices which are simpler than at NLO. The differential cross section in the azimuthal angle 
$\phi=\theta_1-\theta_2-\pi$, with $\theta_i$ being the angles of the two tagged jets, is
\begin{eqnarray}
\frac{d {\hat \sigma}\left(\alpha_s, {\rm Y},p_{1,2}^2\right)}{d \phi}  &=&
\frac{\pi^2 {\bar \alpha}_s^2}{4 \sqrt{p_1^2 p_2^2}} \sum_{n=-\infty}^\infty 
e^{i n \phi} \, {\cal C}_n \left({\rm Y}\right), \nonumber\\
{\cal C}_n \left({\rm Y}\right) &=&
\frac{1}{2 \pi}\int_{-\infty}^\infty \frac{d \nu}{\left(\frac{1}{4}+\nu^2\right)}\left(\frac{p_1^2}{p_2^2}\right)^{i \nu} e^{\chi \left(\left|n\right|,\frac{1}{2}+ i \nu,{\bar \alpha}_s \left(p_1 p_2\right)\right){\rm Y} }, \nonumber\\
\chi \left(n,\gamma,{\bar \alpha}_s\right) &=& 
{\bar \alpha}_s \chi_0\left(n,\gamma\right)
+{\bar \alpha}_s^2 \left(\chi_1\left(n,\gamma\right)
-\frac{\beta_0}{8 N_c} \frac{\chi_0\left(n,\gamma\right)}
{\gamma \left(1-\gamma\right)}\right).\nonumber
\end{eqnarray}
The eigenvalue of the LO kernel is $\chi_0 \left(n,\gamma\right) = 2 \psi \left(1\right) 
- \psi \left(\gamma+ \frac{n}{2}\right) - \psi\left(1-\gamma+\frac{n}{2}\right)$,
with $\psi$ the logarithmic derivative of the Euler function. The action of 
$\hat{K}_1$, in $\overline{\rm MS}$ scheme, can be found in~\cite{Kotikov:2000pm}. 
The full cross section only depends on the $n=0$ component, ${\hat \sigma} =  
\frac{\pi^3 {\bar \alpha}_s^2}{2 \sqrt{p_1^2 p_2^2}} \, {\cal C}_0 \left({\rm Y}\right)$.
The average of the cosine of the 
azimuthal angle times an integer projects out the contribution from each of 
these angular components:
\begin{eqnarray}
\frac{\left<\cos{\left( m \phi \right)}\right>}{\left<\cos{\left( n \phi \right)}\right>} &=& \frac{{\cal C}_m \left({\rm Y}\right)}{{\cal C}_n\left({\rm Y}\right)}
\label{Ratiosformula}
\end{eqnarray}
The normalized differential cross section is
\begin{eqnarray}
\frac{1}{{\hat \sigma}}\frac{d{\hat \sigma}}{d \phi} ~=~
\frac{1}{2 \pi}\sum_{n=-\infty}^\infty 
e^{i n \phi} 
\frac{{\cal C}_n\left({\rm Y}\right)}
     {{\cal C}_0\left({\rm Y}\right)} 
~=~ \frac{1}{2\pi}
\left\{1+2 \sum_{n=1}^\infty \cos{\left(n \phi\right)}
\left<\cos{\left( n \phi \right)}\right>\right\}.
\label{fullangular}
\end{eqnarray}
The BFKL resummation is not stable at NLO~\cite{gavin1,agus05}. In 
the gluon--bremsstrahlung scheme our distributions become unphysical. 
To improve the convergence we impose compatibility with renormalization group evolution 
in the DIS limit~\cite{Ciafaloni:2003rd} for all angular components. 
A good scheme is the angular extension of that discussed in~\cite{agus05}, first 
proposed in~\cite{gavin1}:
\begin{eqnarray}
\omega  &=& {\bar \alpha}_s \left(1+ {\cal A}_n {\bar \alpha}_s\right)
\left\{2 \, \psi \left(1\right) 
- \psi \left(\gamma + \frac{\left|n\right|}{2}+\frac{\omega}{2}+{\cal B}_n {\bar \alpha}_s \right) \right. \\
&&\hspace{-1cm}- \left.\psi \left(1-\gamma + \frac{\left|n\right|}{2}+\frac{\omega}{2}+{\cal B}_n {\bar \alpha}_s \right) \right\} + {\bar \alpha}_s^2 \, \Bigg\{\chi_1 \left(\left|n\right|,\gamma\right)-\frac{\beta_0}{8 N_c} \frac{\chi_0\left(n,\gamma\right)}{\gamma \left(1-\gamma\right)}\nonumber\\
&&\hspace{-1cm}-{\cal A}_n \chi_0\left(\left|n\right|,\gamma\right)\Bigg)
+ \left(\psi'\left(\gamma + \frac{\left|n\right|}{2}\right)
+\psi' \left(1-\gamma + \frac{\left|n\right|}{2}\right) \right)
\left(\frac{\chi_0\left(\left|n\right|,\gamma\right)}{2}+{\cal B}_n\right)
\Bigg\},\nonumber
\end{eqnarray}
where ${\cal A}_n$ and ${\cal B}_n$ are collinear coefficients. After this collinear resummation 
our observables have a good physical behavior and are independent of the renormalization scheme. 

\section{Phenomenology} 

The D$\emptyset$~\cite{Abachi:1996et} collaboration analyzed data for Mueller--Navelet jets at $\sqrt{s} = 630$ and 
1800 GeV. For the angular correlation LO BFKL predictions were first 
obtained in~\cite{DelDuca:1993mn,Stirling:1994zs} and failed to 
describe the data estimating too much decorrelation. An exact fixed NLO analysis 
using JETRAD underestimated the decorrelation, while HERWIG was in agreement with the data.
\begin{figure}[htbp]
  \centering
  \includegraphics[width=7cm]{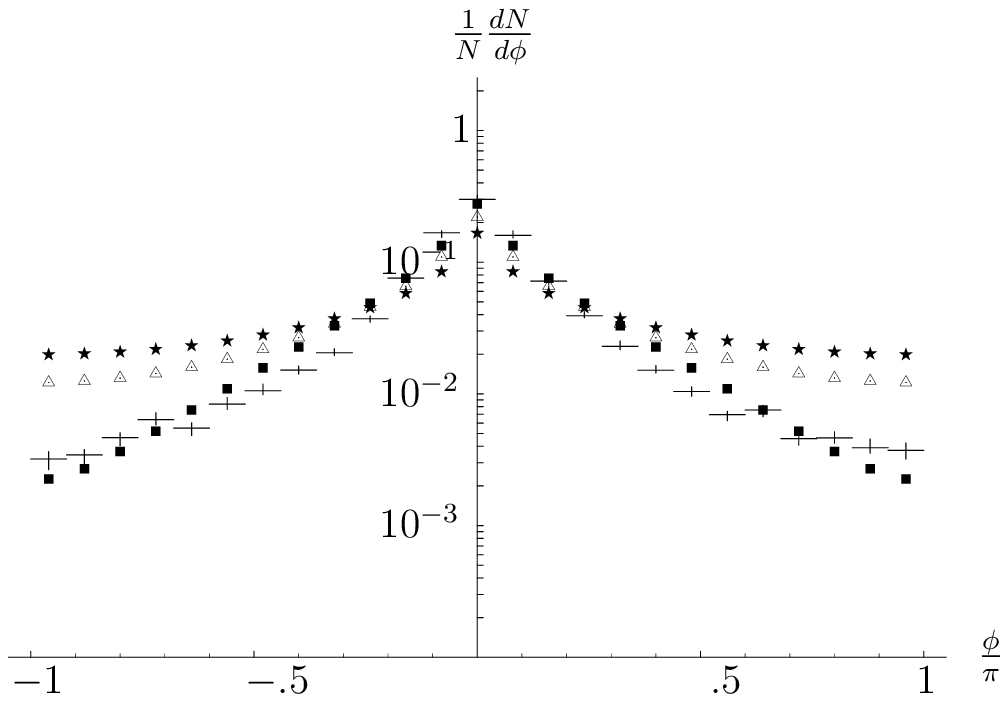}\includegraphics[width=7cm]{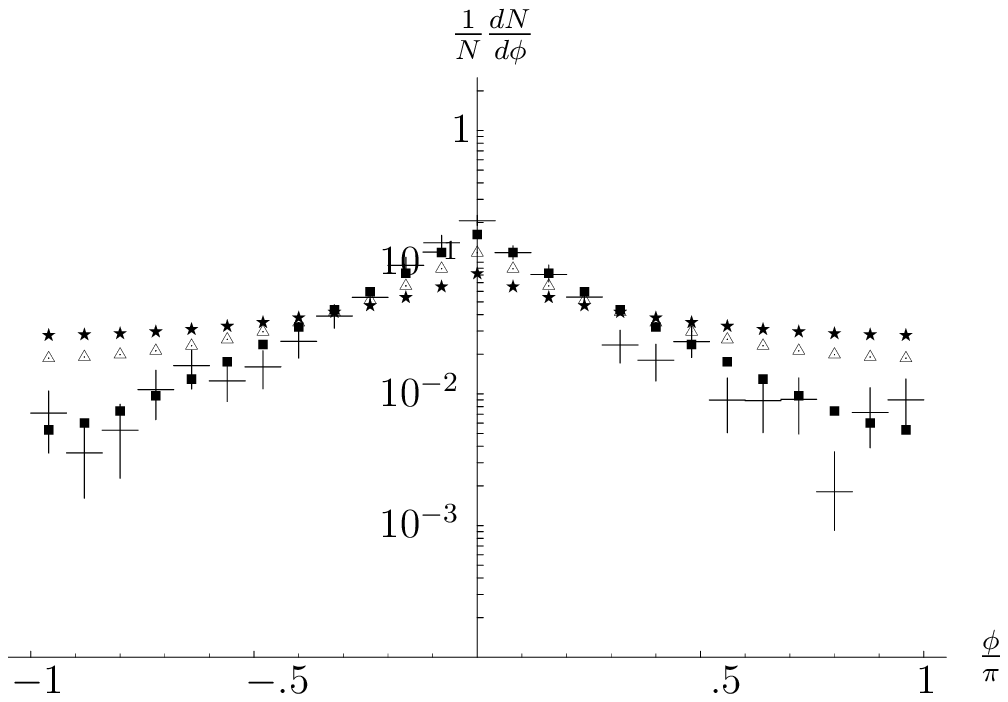}
   \caption{$\frac{1}{N}\frac{dN}{d\phi}$ in a $p\bar{p}$ collider at $\sqrt{s}$=1.8 TeV using a LO (stars), NLO (squares) and resummed (triangles) BFKL kernel. Plots are shown for Y = 3 (left) and Y = 5 (right).}
  \label{fig:tevatrondsigma}
\end{figure}

\begin{figure}[htbp]
  \centering
  \includegraphics[width=7cm]{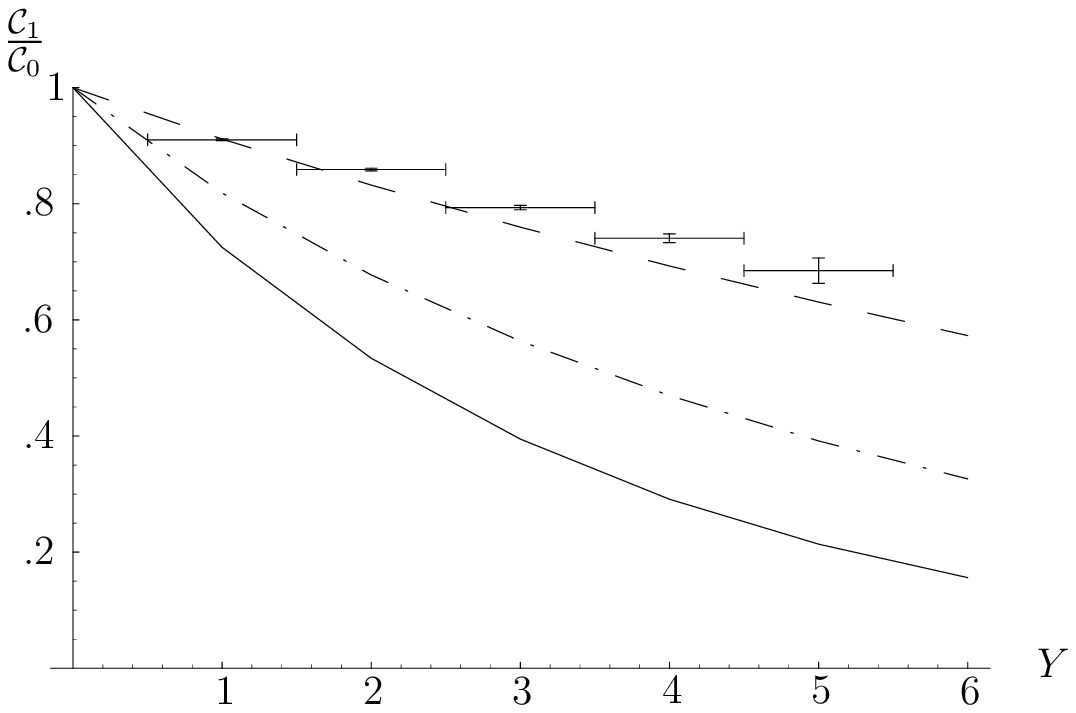}\includegraphics[width=7cm]{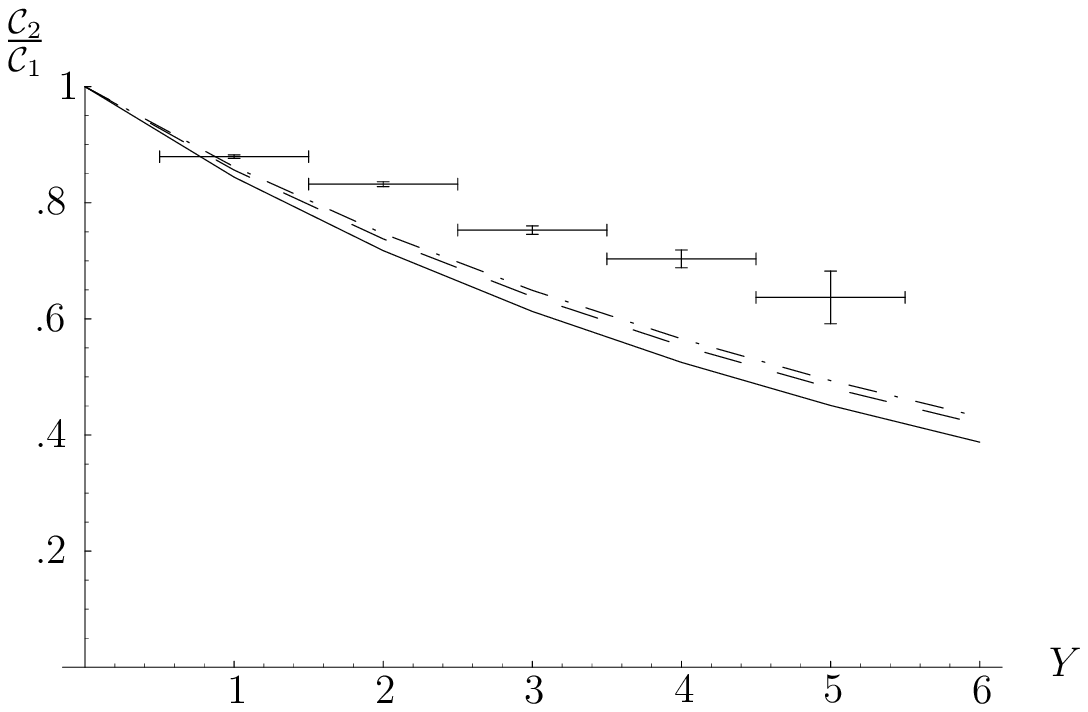}
\caption{Left: $\left<\cos\phi\right> = {\cal C}_1/{\cal C}_0$ and Right: 
$\frac{<\cos2\phi>}{<\cos\phi>} = \frac{{\cal C}_2}{{\cal C}_1}$, at a 
$p\bar{p}$ collider with $\sqrt{s}$ = 1.8 TeV for BFKL at LO (solid) and 
NLO (dashed). The results from the resummation presented in the text are 
shown as well (dash--dotted).}
\label{fig:tevatron1}
\end{figure}

\begin{wrapfigure}{r}{0.5\columnwidth}
\centerline{\includegraphics[width=0.45\columnwidth]{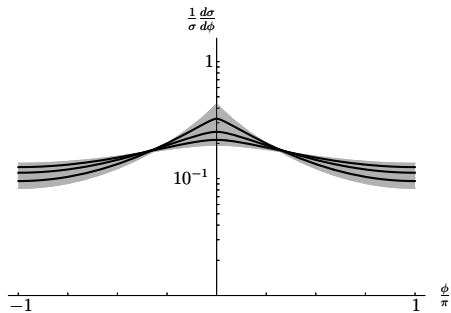}}
\caption{$\frac{1}{\sigma}\frac{d\sigma}{d\phi}$ in our resummation scheme for 
rapidities Y = 7, 9, 11 from top to bottom. The gray band reflects the 
uncertainty in $s_0$ and in the renormalization scale $\mu$.}
\label{fig:lhcdsigma}
\end{wrapfigure}

In Fig.~\ref{fig:tevatron1} we compare the Tevatron data for $\left<\cos\phi\right> = {\cal C}_1/{\cal C}_0$ with our LO, NLO and resummed predictions. For Tevatron's cuts, where the transverse momentum for one 
jet is 20 GeV and for the other 50 GeV, the NLO calculation is instable under renormalization 
scheme changes. The convergence of our observables is poor 
whenever the coefficient associated to zero conformal spin, ${\cal C}_0$, is 
involved. If we eliminate this coefficient by calculating the ratios defined in 
Eq.~(\ref{Ratiosformula}) then the predictions are very stable, see Fig.~\ref{fig:tevatron1}.
The full angular dependence studied at the Tevatron 
by the {D$\emptyset$} collaboration was published 
in~\cite{Abachi:1996et}. In Fig.~\ref{fig:tevatrondsigma} we compare this 
measurement with the predictions obtained in our approach. 
For the differential cross section we also make predictions for the LHC at larger Y in  
Fig.~\ref{fig:lhcdsigma}. 
Our calculation is not exact and we estimated several 
uncertainties, which are represented by gray bands.

\section{Conclusions}

We have presented an analytic study of NLO BFKL corrections in azimuthal angle 
decorrelations for Mueller--Navelet jets at hadron colliders. 
We found that the intercepts for non--zero conformal spins have 
good convergence. The zero conformal spin component needs of a collinear improvement 
to get stable results.
Uncertainties in our study can be reduced using Monte Carlo 
techniques. We compared to the data 
extracted at the Tevatron many years ago. Our results improve with respect to the LO BFKL 
predictions but show too much azimuthal angle decorrelation. The LHC at CERN will have 
larger rapidity differences and will be a very useful tool to investigate the importance 
of BFKL effects in multijet production~\cite{Bartels:2006hg}.

\end{document}